\documentclass[12pt,letterpaper]{article}

\usepackage{latexsym}
\usepackage{graphicx}
\usepackage{indentfirst}
\usepackage{hyperref}
\usepackage{comment}
\usepackage[shortlabels]{enumitem}
\usepackage[english]{babel}
\usepackage{tikz}
\usepackage{esdiff}
\usepackage{amsthm}
\usepackage{amssymb}
\usepackage{epsfig}
\usepackage{psfrag}
\usetikzlibrary{patterns}
\usepackage{fullpage}
\usepackage{times}
\usepackage{amsmath}
\usepackage{amsfonts}
\usepackage{float}

\newcommand{\myparskip}{2pt}
\parskip \myparskip
\newtheorem{theorem}{Theorem}[section]

\newcommand{\eps}{\epsilon}
\def\bA{{\overline{A}}}
\def\bB{{\overline{B}}}
\def\bC{{\overline{C}}}

\newcommand{\cI}{{\mathcal{I} }}

\title{Analysis of the two-for-one swap heuristic for
approximating the maximum independent set in a $k$-polymatroid\footnote{Minor corrections from the version published in Operations Research Letters}
}

\author{Gruia C\u alinescu\thanks{The Research Institute of the University of Bucharest. Work performed while at Illinois Institute of Technology. Email: gcali9999@gmail.com}
\and Adrian Calinescu\thanks{Work performed while at University of Illinois. Email: adriancyay@gmail.com}}
\date{}

\begin{document}
\maketitle

\begin{abstract}
Let $f:2^N \rightarrow \mathbb{Z}_{\geq 0}$ be a polymatroid (an integer-valued
non-decreasing submodular set function with $f(\emptyset) = 0$). 
A $k$-polymatroid satisfies that $f(e) \leq k$ for all $e \in N$.
We call $S \subseteq N$ {\em independent} if $f(S) = \sum_{e \in S} f(e)$ 
and $f(e) > 0$ for all $e \in S$. Such a set was also called a {\em matching}.
Finding a maximum-size independent set in a 
$2$-polymatroid has been studied
and polynomial-time algorithms are known for linearly represented polymatroids.
For $k \geq 3$, the problem is NP-hard, 
and a $((2/k) - \eps)$-approximation is known and 
is obtained by swapping as long as possible a subset of up to
 $(1/\eps)^{\log_{k-1} (2k+1) }$
elements from the current solution by a set with one more element. 

Here we give a simple analysis of
the more particular two-for-one repeated swapping heuristic,
obtaining a tight (weaker) $(2/(k+1))$-approximation.
\end{abstract}

\noindent\textbf{Keywords:} Approximation Algorithm; Matroid; Local Optimization.

\noindent\textbf{2000 MSC:} 05B35; 05B40; 68W05; 68W25; 68W40; 68R05; 90C27.

\section{Introduction}
\label{sec:intro}
A polymatroid $f:2^N \rightarrow \mathbb{Z}_{\geq 0}$ on a ground set $N$ is a
non-decreasing (monotone) integer-valued submodular function
with $f(\emptyset) = 0$.
A function $f$ is monotone iff $f(A) \leq f(B)$ for all $A \subseteq B$.
A function $f$ is submodular iff $f(A) + f(B) \geq f(A\cup B) + f(A\cap
B)$ for all $A, B \subseteq N$.\footnote{Equivalently,
a monotone function $f$ is submodular if
for any $S \subseteq T \subseteq N$, we have 
$\sum_{i \in T \setminus S} (f(S \cup \{i\})-f(S))\geq f(T)-f(S)$.
\cite[Proposition 7]{NWF}}
Polymatroids generalize matroids which have the additional
condition that $f(\{i\}) \leq 1$ for all $i \in N$.

Matroids are versatile mathematical structures that
 generalize numerous combinatorial models, including  
matching \cite{Lawler76, LSV13}, 
online computation \cite{BIK07, BIKK18}, 
diversity maximization \cite{AMT13}
and variants of clustering problems \cite{CLLW16, KKN_11, KKN_15, Swa16}.

A $k$-polymatroid satisfies that $f(e) \leq k$ for all $e \in N$.
We call $S \subseteq N$ {\em independent} if $f(S) = \sum_{e \in S} f(e)$ 
and $f(e) > 0$ for all $e \in S$.
Such a set was also called a {\em matching} by \cite{lovasz1980selecting}.
The {\sc Maximum Independent Set in a $k$-Polymatroid} (or MISkP)
asks for finding a maximum-size independent set in a $k$-polymatroid.
Finding maximum-size sets in downward-closed set systems is one of the most
fundamental combinatorial optimization problems \cite{CCPS98,Schrijver03}.

MISkP was also studied under the name of Matroid $k$-Parity.
MIS2P has been proposed by Lawler \cite{Lawler76} 
and polynomial-time algorithms are known for linearly represented polymatroids
\cite{Orlin08,Lovasz,CLL14,GS86}.
 There are $2$-polymatroids for which
MIS2P is NP-hard \cite{Schrijver03},
 and Jensen and Korte \cite{JK82} have shown that
there exists no algorithm in which the matroid is represented by an independence test oracle (or an oracle polynomially related to an independence test oracle) and which solves the problem in question after a number of calls on the oracle which is bounded by a polynomial in the number of elements of the ground set of the matroid.

Lee, Sviridenko, and Vondr\'{a}k \cite{LSV13} give a polynomial-time approximation scheme
for MIS2P using the local search method:
start with a feasible current solution, and
as long as we obtain a feasible solution, swap
a subset of up to $j$ elements of the current solution by 
a set containing one more element.
The algorithm stops when no such swap is possible.
  Precisely, to get a $(1-\eps)$-approximation, they use a $j$
that is exponential in $1/\eps$.
For MISkP with $k \geq 3$, \cite{LSV13} obtain a $((2/k) - \eps)$-approximation
for this swapping heuristic with 
$j = \lceil (1/\eps)  ^{\log_{k-1} (2k+1)} \rceil $
(\cite{LSV13} only discusses the case where $f(e) = k$ for all $e \in N$,
the $k$-uniform case, but after deleting elements with $f(e)=0$,
one can set $f'(X)=f(X)+\sum_{e \in X}(k-f(e))$,
so that $f'(e) = k$ for all remaining $e$ and
a set is independent with respect to $f$ if and only if it
is independent with respect to $f'$).
Note that $\log_{k-1}(2k+1) \leq 3$ and for large $k$, it is close to $1$.
It is also known that a ratio better than $\Omega(\log k/k)$ is not
possible unless $P=NP$ \cite{HSS06}.

In this paper, we study this local-swap heuristic with $j=1$ 
and obtain a tight approximation ratio of $2/(k+1)$. This is
weaker than the bound of \cite{LSV13}, and MISkP is most
useful when $k$ is small (i.e. MIS3P may help with 
Maximum Planar Subgraph \cite{CFFK98}
and related problems where MIS2P
is used for the best known approximation).  
We have a simple proof for our (weaker) result
which may open the way for proving a 
$( (2/k) - \eps ) $-approximation with
swapping fewer elements, because for fixed $k \geq 3$, $j = O(\log (1/\eps))$ suffices
for the much simpler case of finding a maximum
matching in a $k$-hypergraph \cite{HS89}.
Also, one can further try to apply the method of Cygan \cite{Cygan13}
with the goal of a $((3/(k+1)) - \eps)$-approximation.
Our method is based on
exchange properties of Greene and Magnanti \cite{GM75}
(with special cases in \cite{Gr73,B73},
also Exercise 8.48 in \cite{CCPS98})
and a particular (base) case of the analysis of 
Hurkens and Schrijver \cite{HS89}.

As explained in \cite{LSV13},
MISkP can approximately solve a number of other problems such as 
finding the maximum-size set that is independent in $k$ given matroids
on the same ground set, which we call the
 $k$-Matroid Intersection Problem.
 $2$-Matroid Intersection 
is a classical combinatorial optimization problem that admits
polynomial-time algorithms \cite{E70,Schrijver03}.
Previous to \cite{LSV13},   $k$-Matroid Intersection
had a approximation ratio of $1/(k-1)$ \cite{LRS11}.

We do mention that only unweighted problems were discussed before.
Edmonds \cite{E79} (also in \cite{Schrijver03})
 gives exact polynomial-time algorithms for 
Max-Weight $2$-Matroid Intersection.
For weighted MIS2P in linearly represented matroids
\cite{CGM92,PromelS00} give pseudopolynomial algorithms
and approximation schemes, and Iwata and Kobayashi
\cite{iwata2021weighted} gives a
polynomial-time algorithm. Linhares, Olver, Swamy, and Zenklusen
\cite{LISZ20} gives a $(1/2)$-approximation
for Max-Weight  $3$-Matroid Intersection.

\section{Preliminaries}

In the introduction, a matroid was defined as a polymatroid
which has the additional
condition that $f(\{i\}) \leq 1$ for all $i \in N$.
An alternative definition is:
A {\em matroid} (defined in \cite{CCPS98,Schrijver03})
is an ordered pair $M = (E, \cI)$ satisfying the following conditions:
\begin{enumerate}
\item $E$ is a nonempty finite set.
\item $\cI$ is a nonempty family of subsets of $E$,
 called {\em independent} subsets of $E$,
such that if $B \in \cI$ and $A \subseteq B$, then $A \in \cI$.
\item If $A \in \cI, B \in \cI$,
 and $\vert A\vert < \vert B \vert$,
then there exists some element $x \in B \setminus A$ 
such that $A \cup \{x\} \in \cI$.
\end{enumerate}
To construct $M = (E, \cI)$ from the polymatroid $f$, let $E=N$
and $\cI$ be the sets
that are defined as independent in the polymatroid $f$.
It is known \cite{Schrijver03} and not hard to prove that 
the three conditions above hold, assuming that 
$f(\{i\}) \leq 1$ for all $i \in N$.
To construct a polymatroid from the alternative definition above,
for set $A \subseteq E$, define  the {\em rank} function
$r(A) = \max_{B \subseteq A \mbox{ and }  B \in \cI } |B|$.
This definition implies that $A \in \cI$ if and only if $|A| = r(A)$.
It is not hard to prove (see \cite{Schrijver03}) that $r()$ is a monotone
submodular function with $r(\emptyset) = 0$,
and thus $r$ is a polymatroid. Also,
$r(\{i\}) \leq 1$ for all $i \in E$.

Given a matroid $M = (E, \cI)$, we say that $A \in \cI$
is a maximal independent subset of $M$,
 or a {\em basis} of $M$ if there exists no $x\in E\backslash A$ such that 
$A \cup \{x\} \in \cI$.
It is known that all bases of a matroid have the same size.
We will use:

\begin{theorem}[Theorem 3.3 of \cite{GM75}]
Consider a matroid and let $Q,P$ be two of its bases.
Let $X_1, X_2, \ldots, X_t$ be a partition of $Q$.
Then there is a partition $Y_1, Y_2, \ldots, Y_t$ of $P$ such that for all $i = 1, 2, \ldots, t$, $(Q \setminus X_i) \cup Y_i$ is a basis.
\label{t_exc}
\end{theorem}

Given matroid $M = (E, \cI)$ and 
$Z \subseteq E$, let $X$ be a maximal independent subset of $Z$.
Let $M/Z = (E', \cI')$ where $E' = E \setminus Z$
 and $\cI' = \{ A \subseteq E' \big\vert\ A \cup  X \in \cI \}$.
We call $M/Z$ the {\em contraction} of $M$ by set $Z$ 
(it is known and not hard to prove that the choice of $X$ does not matter).
The rank function $r'$ of $M/Z$ satisfies $r'(Y)  = r(Y \cup Z) - r(Z)$
 (this is not hard to see, and appears in \cite{Schrijver03}).

Helgason \cite{H74}, McDiarmid \cite{M75}, and Lovasz \cite{L77}
have shown that for any polymatroid on set $N$ given by function $f$
there exists a matroid $M = (E,\cI)$
with rank function $r$ with the following properties:
Every element $h \in N$ is associated with an independent set $M_h \subseteq E$,
all the $M_h$ are pairwise disjoint, 
and for all $X \subseteq N$ we have $f(X)= r(\cup_{h \in X} M_h)$.
This exact set of conditions occurs in \cite{Schrijver03} at subsection 44.6b. Although we use this underlying structure of polymatroids for our analysis, at no point is it necessary to run the algorithm.

\section{The analysis of the two-for-one swap heuristic}

We first set up the matroid in which we will apply
Theorem \ref{t_exc}; this is taken from \cite{LSV13} with
slightly  changed (clarified)
notation (for this set up, we had another method based on Theorem \ref{t_exc}
before we discovered \cite{LSV13}).

Let $A \subseteq N$ be our current solution and $B \subseteq N$
be an optimum solution. Assume no feasible addition or two-for-one swap is possible,
from this we will prove that $|A| \geq (2/(k+1)) |B|$.
If $A=\emptyset$, the absence of a feasible addition implies $B=\emptyset$;
assume henceforth that $A \neq \emptyset$.
Let $M$ be the matroid with the rank function $r$
associated to the polymatroid on 
$N$ defined by $f()$ as explained at the end of the previous section.
Let $\bA = \cup_{h \in A} M_h$ and $\bB = \cup_{h \in B} M_h$.
We remove from $M$ all elements outside of $\bA \cup \bB$ as we will not use them for our analysis. 
We have
$r(\bA) = f(A) =  \sum_{h \in A} f(h) = \sum_{h \in A } r(M_h)
= \sum_{h \in A} |M_h| = |\bA|$, where the first equality is from the definition of $M$,
the second equality is from the fact that $A$ is independent in the polymatroid, 
the third equality is again from the properties of $M$, the fourth equality is from 
the fact that each $M_h$ is independent in $M$ (as seen in the properties of $M$),
and the last equality from the fact that
the sets $M_h$ are pairwise disjoint (as seen in the properties of $M$).
We conclude that $\bA$ is independent in $M$.
The same reasoning is applied to obtain that  $\bB$ is independent in $M$.

\begin{figure}
    \centering
\tikzset{
pattern size/.store in=\mcSize, 
pattern size = 5pt,
pattern thickness/.store in=\mcThickness, 
pattern thickness = 0.3pt,
pattern radius/.store in=\mcRadius, 
pattern radius = 1pt}
\makeatletter
\pgfutil@ifundefined{pgf@pattern@name@_q3z7qpwaz}{
\pgfdeclarepatternformonly[\mcThickness,\mcSize]{_q3z7qpwaz}
{\pgfqpoint{0pt}{0pt}}
{\pgfpoint{\mcSize+\mcThickness}{\mcSize+\mcThickness}}
{\pgfpoint{\mcSize}{\mcSize}}
{
\pgfsetcolor{\tikz@pattern@color}
\pgfsetlinewidth{\mcThickness}
\pgfpathmoveto{\pgfqpoint{0pt}{0pt}}
\pgfpathlineto{\pgfpoint{\mcSize+\mcThickness}{\mcSize+\mcThickness}}
\pgfusepath{stroke}
}}
\makeatother
\tikzset{every picture/.style={line width=0.75pt}} 

 
\tikzset{
pattern size/.store in=\mcSize, 
pattern size = 5pt,
pattern thickness/.store in=\mcThickness, 
pattern thickness = 0.3pt,
pattern radius/.store in=\mcRadius, 
pattern radius = 1pt}
\makeatletter
\pgfutil@ifundefined{pgf@pattern@name@_tgrdpi8o9}{
\pgfdeclarepatternformonly[\mcThickness,\mcSize]{_tgrdpi8o9}
{\pgfqpoint{0pt}{0pt}}
{\pgfpoint{\mcSize+\mcThickness}{\mcSize+\mcThickness}}
{\pgfpoint{\mcSize}{\mcSize}}
{
\pgfsetcolor{\tikz@pattern@color}
\pgfsetlinewidth{\mcThickness}
\pgfpathmoveto{\pgfqpoint{0pt}{0pt}}
\pgfpathlineto{\pgfpoint{\mcSize+\mcThickness}{\mcSize+\mcThickness}}
\pgfusepath{stroke}
}}
\makeatother
\tikzset{every picture/.style={line width=0.75pt}} 

\begin{tikzpicture}[x=0.75pt,y=0.75pt,yscale=-1,xscale=1]

\draw  [color={rgb, 255:red, 31; green, 87; blue, 156 }  ,draw opacity=1 ][line width=2.25]  (102,83.2) .. controls (102,80.88) and (103.88,79) .. (106.2,79) -- (287,79) .. controls (289.32,79) and (291.2,80.88) .. (291.2,83.2) -- (291.2,232.8) .. controls (291.2,235.12) and (289.32,237) .. (287,237) -- (106.2,237) .. controls (103.88,237) and (102,235.12) .. (102,232.8) -- cycle ;
\draw  [line width=1.5]  (100,32.4) .. controls (100,26.1) and (105.1,21) .. (111.4,21) -- (213.8,21) .. controls (220.1,21) and (225.2,26.1) .. (225.2,32.4) -- (225.2,66.6) .. controls (225.2,72.9) and (220.1,78) .. (213.8,78) -- (111.4,78) .. controls (105.1,78) and (100,72.9) .. (100,66.6) -- cycle ;
\draw  [color={rgb, 255:red, 31; green, 87; blue, 156 }  ,draw opacity=1 ][pattern=_tgrdpi8o9,pattern size=6pt,pattern thickness=0.75pt,pattern radius=0pt, pattern color={rgb, 255:red, 31; green, 87; blue, 156}][line width=2.25]  (200.2,31.4) .. controls (200.2,25.1) and (205.3,20) .. (211.6,20) -- (281.8,20) .. controls (288.1,20) and (293.2,25.1) .. (293.2,31.4) -- (293.2,65.6) .. controls (293.2,71.9) and (288.1,77) .. (281.8,77) -- (211.6,77) .. controls (205.3,77) and (200.2,71.9) .. (200.2,65.6) -- cycle ;
\draw  [color={rgb, 255:red, 208; green, 2; blue, 27 }  ,draw opacity=1 ][line width=3]  (168,31.4) .. controls (168,25.1) and (173.1,20) .. (179.4,20) -- (281.8,20) .. controls (288.1,20) and (293.2,25.1) .. (293.2,31.4) -- (293.2,65.6) .. controls (293.2,71.9) and (288.1,77) .. (281.8,77) -- (179.4,77) .. controls (173.1,77) and (168,71.9) .. (168,65.6) -- cycle ;

\draw (134,38) node [anchor=north west][inner sep=0.75pt]   [align=left] {$\displaystyle \overline{A}$};
\draw (186,148) node [anchor=north west][inner sep=0.75pt]  [color={rgb, 255:red, 31; green, 87; blue, 156 }  ,opacity=1 ] [align=left] {$\displaystyle \overline{B} '$};
\draw (303,33) node [anchor=north west][inner sep=0.75pt]  [color={rgb, 255:red, 31; green, 87; blue, 156 }  ,opacity=1 ] [align=left] {$\displaystyle \overline{B} ''$};
\draw (180,36) node [anchor=north west][inner sep=0.75pt]  [color={rgb, 255:red, 208; green, 2; blue, 27 }  ,opacity=1 ] [align=left] {$\displaystyle \overline{C}$};

\end{tikzpicture}

    \caption{Illustrating the definitions of $\bA$ - the black rounded rectangle, 
    $ \bB'$ - the empty blue rounded rectangle, $\bB''$ - the filled blue rounded rectangle,
    and $\bC$ - the red rounded rectangle. Keep in mind that $\bB = \bB' \cup \bB''$
    and  $\bB' \cap \bB'' = \emptyset$.}
    \label{f_setup}
\end{figure}
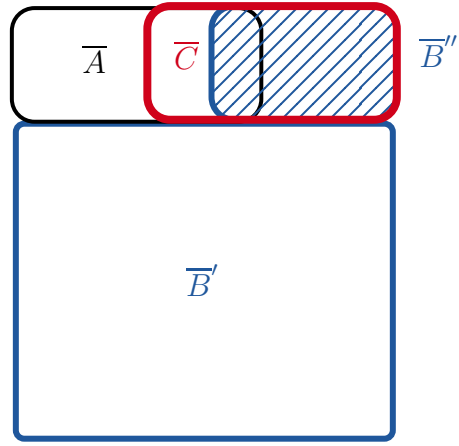

Since $\bA$ is independent in $M$,
we have a maximal independent set $\bar{I}$ in $M$ that contains $\bA$.
Define $\bB '=\bar{I}\backslash \bA$; as the ground set of $M$ is $\bA \cup \bB$,
we have that $\bB' \subseteq \bB$.
From here our analysis diverges from \cite{LSV13}.
Contract $\bB'$ in $M$ to obtain matroid $M'$.
Let $\bB'' := \bB \setminus \bB'$.
As $\bB'$ is independent in $M$, by the definition of contraction,
the independence of set $\bA$ in $ M'$ corresponds to 
the independence of $\bA \cup \bB'$ in $ M$.
Also, the independence of $\bB''$ in $ M'$ corresponds to
the independence of $\bB'' \cup \bB'=\bB$ in $M$.
As $\bA \cup \bB'$ is a basis in $M$, $\bA$ is a basis in $M'$,
and similarly $\bB$ being independent in $M$ implies
that $\bB''$ is an independent set in $M'$.
As such $\bB ''\subseteq \bC$ for some basis $\bC$ in $M'$.
See Figure \ref{f_setup} for an illustration.


\begin{figure}[htbp]
    \centering

\tikzset{every picture/.style={line width=0.75pt}} 

\begin{tikzpicture}[x=0.75pt,y=0.75pt,yscale=-1,xscale=1]

\draw  [line width=2.25]  (106,20.6) -- (187.2,20.6) -- (187.2,204.6) -- (106,204.6) -- cycle ;
\draw    (106,50.6) -- (187.2,50.6) ;
\draw    (105,79.6) -- (152.2,79.6) -- (186.2,79.6) ;
\draw    (106,110.6) -- (187.2,110.6) ;
\draw    (105.8,159.6) -- (147,159.6) -- (173,159.6) -- (187,159.6) ;
\draw  [line width=2.25]  (262,20.6) -- (343.2,20.6) -- (343.2,204.6) -- (262,204.6) -- cycle ;
\draw [color={rgb, 255:red, 0; green, 0; blue, 0 }  ,draw opacity=1 ]   (263,49.6) -- (344.2,49.6) ;
\draw    (262,78.6) -- (309.2,78.6) -- (343.2,78.6) ;
\draw    (263,109.6) -- (344.2,109.6) ;
\draw    (262,158.6) -- (303.2,158.6) -- (323.2,158.6) -- (343.2,158.6) ;
\draw   (106,204.6) -- (343.2,204.6) -- (343.2,341.6) -- (106,341.6) -- cycle ;
\draw  [fill={rgb, 255:red, 208; green, 2; blue, 27 }  ,fill opacity=1 ] (342.89,67.99) .. controls (340.78,68.58) and (338.53,68.9) .. (336.2,68.9) .. controls (324.05,68.9) and (314.2,60.26) .. (314.2,49.6) .. controls (314.2,38.94) and (324.05,30.3) .. (336.2,30.3) .. controls (338.53,30.3) and (340.78,30.62) .. (342.89,31.21) -- cycle ;
\draw  [fill={rgb, 255:red, 65; green, 117; blue, 5 }  ,fill opacity=1 ] (266.2,222.6) .. controls (266.2,201.06) and (273.59,183.6) .. (282.7,183.6) .. controls (291.81,183.6) and (299.2,201.06) .. (299.2,222.6) .. controls (299.2,244.14) and (291.81,261.6) .. (282.7,261.6) .. controls (273.59,261.6) and (266.2,244.14) .. (266.2,222.6) -- cycle ;
\draw  [fill={rgb, 255:red, 74; green, 144; blue, 226 }  ,fill opacity=1 ] (310.2,223.6) .. controls (310.2,202.06) and (317.59,184.6) .. (326.7,184.6) .. controls (335.81,184.6) and (343.2,202.06) .. (343.2,223.6) .. controls (343.2,245.14) and (335.81,262.6) .. (326.7,262.6) .. controls (317.59,262.6) and (310.2,245.14) .. (310.2,223.6) -- cycle ;
\draw  [fill={rgb, 255:red, 208; green, 2; blue, 27 }  ,fill opacity=1 ] (341.89,322.99) .. controls (339.78,323.58) and (337.53,323.9) .. (335.2,323.9) .. controls (323.05,323.9) and (313.2,315.26) .. (313.2,304.6) .. controls (313.2,293.94) and (323.05,285.3) .. (335.2,285.3) .. controls (337.53,285.3) and (339.78,285.62) .. (341.89,286.21) -- cycle ;
\draw  [fill={rgb, 255:red, 189; green, 16; blue, 224 }  ,fill opacity=1 ] (130.2,270.6) .. controls (130.2,249.06) and (137.59,231.6) .. (146.7,231.6) .. controls (155.81,231.6) and (163.2,249.06) .. (163.2,270.6) .. controls (163.2,292.14) and (155.81,309.6) .. (146.7,309.6) .. controls (137.59,309.6) and (130.2,292.14) .. (130.2,270.6) -- cycle ;
\draw   (454,41.1) .. controls (454,35.52) and (458.52,31) .. (464.1,31) .. controls (469.68,31) and (474.2,35.52) .. (474.2,41.1) .. controls (474.2,46.68) and (469.68,51.2) .. (464.1,51.2) .. controls (458.52,51.2) and (454,46.68) .. (454,41.1) -- cycle ;
\draw   (455,192.1) .. controls (455,186.52) and (459.52,182) .. (465.1,182) .. controls (470.68,182) and (475.2,186.52) .. (475.2,192.1) .. controls (475.2,197.68) and (470.68,202.2) .. (465.1,202.2) .. controls (459.52,202.2) and (455,197.68) .. (455,192.1) -- cycle ;
\draw   (454,81.1) .. controls (454,75.52) and (458.52,71) .. (464.1,71) .. controls (469.68,71) and (474.2,75.52) .. (474.2,81.1) .. controls (474.2,86.68) and (469.68,91.2) .. (464.1,91.2) .. controls (458.52,91.2) and (454,86.68) .. (454,81.1) -- cycle ;
\draw  [fill={rgb, 255:red, 208; green, 2; blue, 27 }  ,fill opacity=1 ] (560,51.1) .. controls (560,45.52) and (564.52,41) .. (570.1,41) .. controls (575.68,41) and (580.2,45.52) .. (580.2,51.1) .. controls (580.2,56.68) and (575.68,61.2) .. (570.1,61.2) .. controls (564.52,61.2) and (560,56.68) .. (560,51.1) -- cycle ;
\draw  [fill={rgb, 255:red, 74; green, 144; blue, 226 }  ,fill opacity=1 ] (562,170.1) .. controls (562,164.52) and (566.52,160) .. (572.1,160) .. controls (577.68,160) and (582.2,164.52) .. (582.2,170.1) .. controls (582.2,175.68) and (577.68,180.2) .. (572.1,180.2) .. controls (566.52,180.2) and (562,175.68) .. (562,170.1) -- cycle ;
\draw  [fill={rgb, 255:red, 65; green, 117; blue, 5 }  ,fill opacity=1 ] (562,226.1) .. controls (562,220.52) and (566.52,216) .. (572.1,216) .. controls (577.68,216) and (582.2,220.52) .. (582.2,226.1) .. controls (582.2,231.68) and (577.68,236.2) .. (572.1,236.2) .. controls (566.52,236.2) and (562,231.68) .. (562,226.1) -- cycle ;
\draw  [fill={rgb, 255:red, 189; green, 16; blue, 224 }  ,fill opacity=1 ] (562,278.1) .. controls (562,272.52) and (566.52,268) .. (572.1,268) .. controls (577.68,268) and (582.2,272.52) .. (582.2,278.1) .. controls (582.2,283.68) and (577.68,288.2) .. (572.1,288.2) .. controls (566.52,288.2) and (562,283.68) .. (562,278.1) -- cycle ;
\draw    (475.2,192.1) -- (562,170.1) ;
\draw    (475.2,192.1) -- (562,226.1) ;
\draw    (474.2,81.1) -- (560,51.1) ;
\draw    (474.2,41.1) -- (560,51.1) ;

\draw (129,26) node [anchor=north west][inner sep=0.75pt]   [align=left] {$\displaystyle M_{h_{1}}$};
\draw (127,55) node [anchor=north west][inner sep=0.75pt]   [align=left] {$M_{h_{2}}$};
\draw (128,87) node [anchor=north west][inner sep=0.75pt]   [align=left] {$M_{h_{3}}$};
\draw (283,27) node [anchor=north west][inner sep=0.75pt]   [align=left] {$\displaystyle Y_{1}$};
\draw (286,55) node [anchor=north west][inner sep=0.75pt]   [align=left] {$\displaystyle Y_{2}$};
\draw (293,84) node [anchor=north west][inner sep=0.75pt]   [align=left] {$\displaystyle Y_{3}$};
\draw (127.6,179.6) node [anchor=north west][inner sep=0.75pt]   [align=left] {$M_{h_{a}}$};
\draw (296,178) node [anchor=north west][inner sep=0.75pt]   [align=left] {$\displaystyle Y_{a}$};
\draw (213,264) node [anchor=north west][inner sep=0.75pt]  [font=\large] [align=left] {$\displaystyle \overline{B} '$};
\draw (353,41) node [anchor=north west][inner sep=0.75pt]  [color={rgb, 255:red, 208; green, 2; blue, 27 }  ,opacity=1 ] [align=left] {$\displaystyle e_{1}$};
\draw (354,296) node [anchor=north west][inner sep=0.75pt]  [color={rgb, 255:red, 208; green, 2; blue, 27 }  ,opacity=1 ] [align=left] {$\displaystyle e_{1}$};
\draw (349,215) node [anchor=north west][inner sep=0.75pt]  [color={rgb, 255:red, 74; green, 144; blue, 226 }  ,opacity=1 ] [align=left] {$\displaystyle e_{2}$};
\draw (246,221) node [anchor=north west][inner sep=0.75pt]  [color={rgb, 255:red, 65; green, 117; blue, 5 }  ,opacity=1 ] [align=left] {$\displaystyle e_{3}$};
\draw (165.2,273.6) node [anchor=north west][inner sep=0.75pt]  [color={rgb, 255:red, 189; green, 16; blue, 224 }  ,opacity=1 ] [align=left] {$\displaystyle e_{4}$};
\draw (599.2,270.6) node [anchor=north west][inner sep=0.75pt]  [color={rgb, 255:red, 189; green, 16; blue, 224 }  ,opacity=1 ] [align=left] {$\displaystyle e_{4}$};
\draw (596,216) node [anchor=north west][inner sep=0.75pt]  [color={rgb, 255:red, 65; green, 117; blue, 5 }  ,opacity=1 ] [align=left] {$\displaystyle e_{3}$};
\draw (596,164) node [anchor=north west][inner sep=0.75pt]  [color={rgb, 255:red, 74; green, 144; blue, 226 }  ,opacity=1 ] [align=left] {$\displaystyle e_{2}$};
\draw (593,37) node [anchor=north west][inner sep=0.75pt]  [color={rgb, 255:red, 208; green, 2; blue, 27 }  ,opacity=1 ] [align=left] {$\displaystyle e_{1}$};
\draw (420,34) node [anchor=north west][inner sep=0.75pt]   [align=left] {$\displaystyle h_{1}$};
\draw (423,72) node [anchor=north west][inner sep=0.75pt]   [align=left] {$\displaystyle h_{2}$};
\draw (429,183) node [anchor=north west][inner sep=0.75pt]   [align=left] {$\displaystyle h_{a}$};
\draw (153.5,119.5) node [anchor=north west][inner sep=0.75pt]  [font=\normalsize,rotate=-90] [align=left] {{\Huge {\fontfamily{ptm}\selectfont \textbf{...}}}};
\draw (307.5,117.5) node [anchor=north west][inner sep=0.75pt]  [font=\normalsize,rotate=-90] [align=left] {{\Huge {\fontfamily{ptm}\selectfont \textbf{...}}}};
\draw (467.5,111.5) node [anchor=north west][inner sep=0.75pt]  [font=\normalsize,rotate=-90] [align=left] {{\Huge {\fontfamily{ptm}\selectfont \textbf{. . .}}}};
\draw (571.5,86.5) node [anchor=north west][inner sep=0.75pt]  [font=\normalsize,rotate=-90] [align=left] {{\Huge {\fontfamily{ptm}\selectfont \textbf{. . .}}}};
\draw (57,97) node [anchor=north west][inner sep=0.75pt]  [font=\large] [align=left] {$\displaystyle \overline{A}$};
\draw (357,96) node [anchor=north west][inner sep=0.75pt]  [font=\large] [align=left] {$\displaystyle \overline{C}$};

\end{tikzpicture}

    \caption{Illustrating the construction of the bipartite graph $G$. Here we pretend that
    $\bA$ and $\bC$ are disjoint (the proof does not use this).  
    A part of $G$ is shown on the right. In the depicted scenario, $e_4$ can be added to
    $A$ and the result is an independent set. Also, 
    $\left( A \setminus \{h_a\} \right) \cup \{e_2, e_3\}$ is an independent set.
    }
    \label{f_bigraph}
\end{figure}
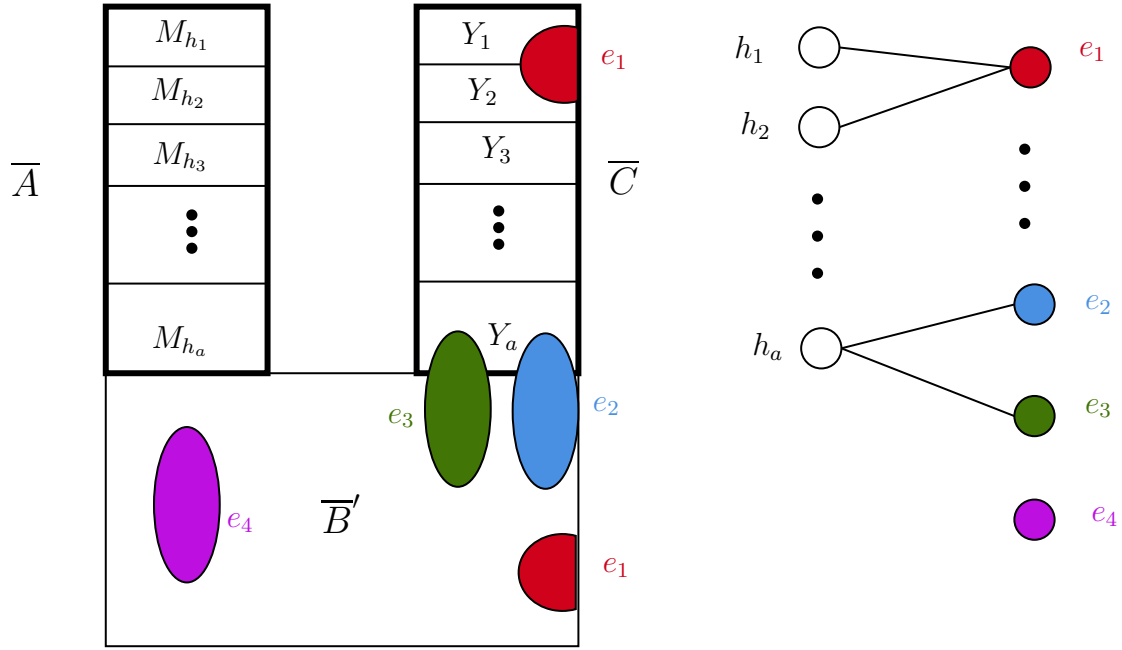

Let $a = |A|$. 
With $A = \{ h_1, h_2, \ldots, h_a\}$, we have that 
$M_{h_1}, M_{h_2}, \ldots, M_{h_a}$ is a partition
of $\bA$. 
Use Theorem \ref{t_exc} to obtain a partition $Y_1, Y_2, \ldots, Y_a$
of $\bC$ such that, for all $i \in \{1,2,\ldots, a\}$, we have
that $\bA \setminus M_{h_i} \cup Y_i$ is a basis in $M'$.
As all bases have the same size, $|Y_i| \geq |M_{h_i}|$ for every $i$.
Since $\sum_i |Y_i| = |\bC| = |\bA| = \sum_i |M_{h_i}|$,
equality holds for every $i$; consequently,
$Y_i \cap (\bA \setminus M_{h_i}) = \emptyset$.
Construct a bipartite graph $G$ with parts $A$ and $B$,
with an exception: for each element in both $A$ and $B$, duplicate it
and put it in both parts of $G$. In $G$,
we put an edge between $h_i \in A$ and $e_j \in B$ if
$M_{e_j} \cap Y_i \neq \emptyset$. See Figure \ref{f_bigraph} for an illustration.
Call a vertex of degree $1$ in $G$ a singleton.

If $h_i \in A \cap B$, then $M_{h_i} \subseteq \bB'' \subseteq \bC$.
The preceding disjointness and equal sizes imply $Y_i=M_{h_i}$.
Thus the two copies of $h_i$ form an isolated edge in $G$.

If a vertex of $e \in B$ has degree $0$ in $G$, then $M_e \cap Y_i = \emptyset$
for all $h_i \in A$, which implies that $M_e \subseteq \bB'$,
and this implies that $A \cup \{e\}$ is independent in
the polymatroid on $N$ given by $f()$.
This cannot happen when the two-for-one heuristic stops.
We conclude that every vertex of $B$ has degree at least one in $G$.

We claim that we cannot have a vertex of $A$ adjacent in $G$ to two
singletons in $B$. Indeed, assume that $h_i \in A$ is adjacent to 
$e$ and $f$ distinct singleton vertices of $B$. 
Then $A' := A \setminus \{h_i\} \cup \{e,f\}$
has size $|A|+1$, since the isolated-edge property gives $e,f \notin A$,
and is an independent set, as we argue next. $e$ being a singleton implies
that all elements of $M_e \cap \bB''$  are in $Y_i$.
Same reasoning applies to $f$. Therefore 
$\cup_{g \in A'} M_g \subseteq 
\left( \bA \setminus M_{h_i} \cup Y_i \right) 
 \cup \left( M_e \cap \bB' \right) 
 \cup \left( M_f \cap \bB' \right)  \subseteq  
 \left( \bA \setminus M_{h_i} \cup Y_i \right)  \cup \bB'$.
As $\bA \setminus M_{h_i} \cup Y_i$ is a basis in $M'$, we have that
$\left( \bA \setminus M_{h_i} \cup Y_i \right)  \cup \bB'$
is independent in $M$, and from this we deduce that
$\cup_{g \in A'} M_g$ is independent in $M$, which implies that
$A'$ is independent in the polymatroid on $N$ given by $f()$.
This cannot happen when the two-for-one heuristic stops.

Now we compare $|A|$ with $|B|$ using the arguments from the base case 
(Case 2) in the proof of Theorem 1 of \cite{HS89}. 
In $G$, the degree of a vertex $h_i \in A$ is at most $k$,
as $|Y_i| = |M_{h_i}| = f(h_i) \leq k$ and $M_{e_j}$ is disjoint from $M_{e_l}$,
for $e_j \neq e_l$, $e_j, e_l \in B$.
Partition $B$ into $S$ and $D$, where $S$ is the set of singletons in $B$,
 and $D$ are the remaining vertices of $G$ that are in $B$.
All the vertices in $D$ have degree at least two, and thus
$k |A| \geq |E(G)| \geq |S| + 2 |D|$.
From the previous paragraph, $|A| \geq |S|$. Add these two last
equations and obtain that $(k+1)|A| \geq 2 |B|$, which 
implies the ratio of at least
$2/(k+1)$ for the two-for-one swap heuristic.

{\bf Note.}  It may be more intuitive to contract $\bA \cap \bB$
early in the proof, but readjusting the definitions will make this proof longer.

\begin{figure}[htbp]
\begin{center}\leavevmode%
\scalebox{0.57}{
  \includegraphics{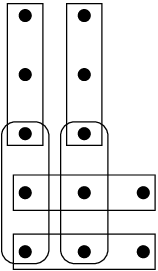}
}
\end{center}
\caption{An example where  the two-for-one swap heuristic
has approximation ratio of exactly $2/(k+1)$, for $k=3$.
The heuristic's output may consists of the two rounded rectangles, while
the four sharp cornered rectangles make an optimum solution.
}
\label{f_tight}
\end{figure}

{\bf Tightness.}
\cite{HS89} prove tight bounds and from their proof
we can immediately obtain that the approximation algorithm
of the two-for-one swap heuristic is exactly $2/(k+1)$.
For $k=1$, every maximal independent set is optimum; assume $k \geq 2$ below.
Assume that $N$ consists of collection of subsets
of $k$ points from some universe $U$,
and $f(A) = | \cup_{e \in A} e |$. It is well known
(and easy to show) that $f()$ is increasing submodular
and so we do have a MISkP instance.
In this setting, a set $A \subseteq N$
 is independent if and only if the elements of $A$ are disjoint.
(this is the  $k$-Set Packing problem \cite{HS89,Cygan13,HSS06}).
Let $e_1, e_2, \ldots, e_{k+1}$ be $k+1$ disjoint sets of $U$
with $k$ elements each, and let $e_{k+2}$ have one element from
each of $e_1, e_3, e_4, \ldots, e_{k+1}$ and $e_{k+3}$
be disjoint from $e_{k+2}$ and have one element from 
each of $e_2, e_3, e_4, \ldots, e_{k+1}$.
The heuristic can stop with $\{e_{k+2},e_{k+3} \}$ as the output
while a solution with $k+1$ elements of $N$ exists: 
$\{e_1, \ldots, e_{k+1} \}$.
This example for $k = 3$ is given in Figure \ref{f_tight}.

\paragraph{Method Limitations}
We are not able to construct this bipartite graph for larger swaps
and general matroids (or linear matroids). 
The method does work for strongly base orderable matroids 
\cite{Brualdi70,Ingleton77,FNSW11,Soto14,DL18} 
and gives the better dependence  of \cite{HS89}
of the approximation ratio as function of the size of the swap.
Constructing such a bipartite graph  for 
strongly base orderable matroids 
was done in Feldman, Naor, Schwartz, and Ward \cite{FNSW11},
but we could not find an explicit dependence of the approximation ratio 
as a function of the size of the swap in their paper.
Note that the matroid being strongly base orderable or not
does not affect the algorithm, just its proven performance guarantee.

\paragraph{Acknowledgment}
Adrian thanks Chandra Chekuri for providing useful references.
Gruia thanks Bahareh Kudarzi for discussions regarding \cite{HS89}.
ChatGPT was used to identify the minor errors that we repair here.

\bibliographystyle{plain}
\bibliography{all}
\label{sec:biblio}

\end{document}